\documentclass{PoS}
\usepackage{url}
\newcommand\apj{The Astrophysical Journal}

\newcommand{\source}{1ES~1218+304~}
\newcommand{\lat}{\textit{Fermi}-LAT~}
\newcommand{\xrt}{\textit{Swift}-XRT~}
\newcommand{\bat}{\textit{Swift}-BAT~}

\newcommand{\veritas}{VERITAS~}

\title{VERITAS Observations of 1ES~1218+304 during the 2019 VHE High State}

\ShortTitle{VERITAS Obseravtions of 1ES~1218+304}

\author{\speaker{Stephan O'Brien}\thanks{For the VERITAS Collaboration}\\
        Department of Physics, McGill University, Montreal, Canada\\
        E-mail: \email{stephan.obrien@mcgill.ca}\\
        }

\abstract{\source is a moderate-redshift (z = 0.182) high-frequency-peaked BL Lac object (HBL). Detected by both the MAGIC and VERITAS gamma-ray observatories, \source is frequently monitored by VERITAS as part of its long-term blazar monitoring program. On the 3rd of January 2019, during a regularly scheduled blazar snapshot, VERITAS observed \source to have an elevated VHE flux state which continued through January 5 (ATEL \#12360). In addition to VERITAS observations, MAGIC also detected elevated VHE emission (ATEL \#12354) and \textit{Swift}-BAT/XRT and Tuorla observations, in X-ray and optical respectively, showed increased multiwavelength activity at the time of the VERITAS observations. In this work, VERITAS observations of \source taken as part of a target of opportunity campaign during this elevated state, are presented.}

\FullConference{36th International Cosmic Ray Conference -ICRC2019-\\
		July 24th - August 1st, 2019\\
		Madison, WI, U.S.A.}

\begin{document}

\section{Introduction}

\source is a high-frequency-peaked BL Lac Object located at a redshift of z = 0.182.
It was discovered as a VHE emitter by the MAGIC collaboration \cite{magicdetection} during an observation campaign in 2005.
During this campaign no significant flux variability was observed on the time-scales of days.
VERITAS has also detected \source, and during a 2008-2009 observation campaign it was observed to have a mean flux of 7\% Crab  \cite{veritas_1es1218}.
Between the nights of 25th January 2009 and 5th February 2009, day-scale variability was observed with the flux reaching a peak of $\sim20\%$ Crab on the night of 30th January 2009.

\par During late 2018, \source entered an active state across multiple wavelengths, with enhanced activity at both X-ray\footnote{\bat- \url{https://swift.gsfc.nasa.gov/results/transients/weak/QSOB1218p304/}} \footnote{\xrt- \url{https://www.swift.psu.edu/monitoring/source.php?source=1ES1218+304} } and optical energies \footnote{Tuorla - \url{http://users.utu.fi/kani/1m/1ES\_1218+304.html}}.
During a regularly scheduled monitoring observation on 3rd January 2019, VERITAS observed an enhanced TeV level of $>$20\% from \source \cite{veritasAtel}.
In response to this elevated state, target of opportunity (ToO) observations were taken the following night by VERITAS.
MAGIC also reported an elevated state during this period \cite{magicAtel}.
VERITAS continued to monitor \source at a high-cadence until 3rd April 2019.
During this period a historic peak in the X-ray count rate was observed by \xrt \cite{xrtAtel}.

In these proceedings, the gamma-ray data taken by VERITAS and \lat from the 2018-2019 season are presented. In Section \ref{sec:obs} the observations taken by VERITAS and the \lat data are described.
In Section \ref{sec:flux} the gamma-ray light curves of \source are discussed and a search for flux variability is performed.
In Section \ref{sec:spec} the gamma-ray spectral energy distribution (SED) for the time-averaged data set is obtained. In addition the SED for an exceptional nightly scale flare is examined.
Finally in Section \ref{sec:con}, the results are summarized.

\section{Observations}
\label{sec:obs}

\par VERITAS (\textbf{V}ery \textbf{E}nergetic \textbf{R}adiation \textbf{I}maging \textbf{T}elescope \textbf{A}rray \textbf{S}ystem) is a gound-based gamma-ray detector, sensitive to gamma rays in the $100\mathrm{~GeV } \mathrm{-} > 30 \mathrm{~TeV}$ energy range.
Located at the Fred Lawrence Whipple Observatory (FLWO) in southern Arizona USA (31 40N, 110 57W,  1.3km a.s.l.), the VERITAS array consists of four 12-m imaging atmospheric-Cherenkov telescopes.
Each telescope has a Davies-Cotton-design segmented mirror dish with 345 facets.
At the focal plane of each dish, lies a a 499 PMT camera, which has a total field of view of $3.5^{\circ}$.
The 68\% containment radius for a 1 TeV photon is $<0.1^{\circ}$, and the pointing accuracy is $<50''$.
In its current configuration VERITAS can detect a source with flux 1\% that of the Crab Nebula in $\sim25$ hours of observations and has an energy resolution of 15-25\%.
For full details of VERITAS and its performance see \cite{VERITAS_Performance}.

\par
\begin{figure}
    \centering
    \includegraphics[width=0.75\textwidth]{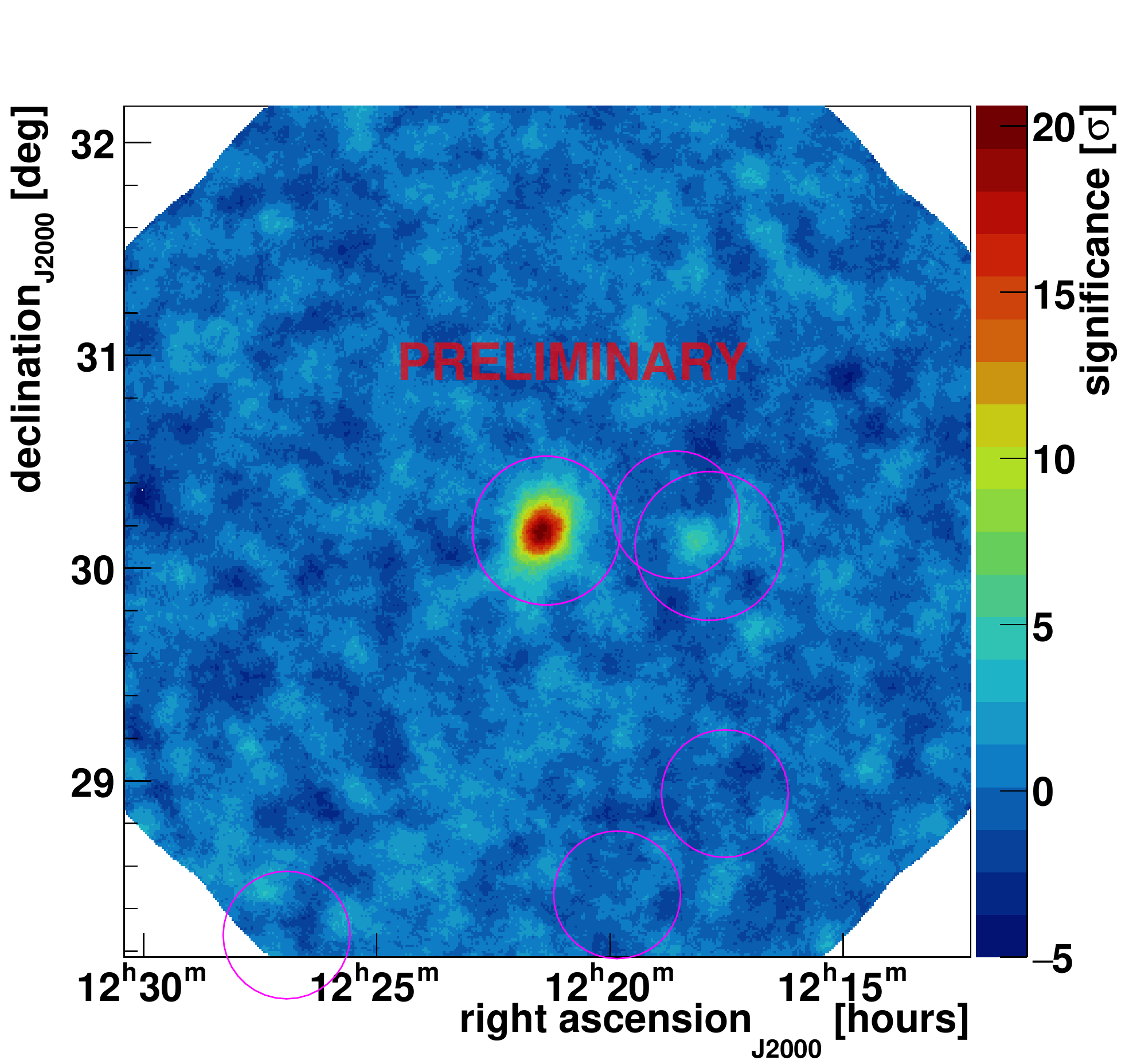}
    \caption{
    Correlated significance sky map for the region of interest.
    The magenta circles indicate regions excluded from the background estimation.
    }
    \label{fig:skymap}
\end{figure}

 VERITAS data taken between 58461 and 58600 MJD (12th December 2018 and  27th April 2019) were analyzed using the two standard VERITAS analysis packages and excellent agreement was found.
Boosted decision tree optimized a priori for soft-spectrum point-source analysis, were used in gamma/hadron separation \cite{bdt_cuts}.
In total the data presented here corresponds to 12.2 hours of dead-time-corrected exposure.
This results in the detection of $556.54$ excess gamma-ray-like events (1087 \texttt{ON}-source, 3789 \texttt{OFF}-source, mean On/Off normalization of 0.14), giving a detection of 19.4 standard deviations above the background ($\sigma$)\cite{liandma}.
The correlated significance sky map of the region is shown in Figure \ref{fig:skymap}, where \source is clearly visible at the centre.
In performing background estimation, $0.35^{\circ}$ radius circles around bright stars are excluded from background estimation.
In addition, there is a known near-by VHE emitter, 1ES~1215+303\footnote{\url{http://tevcat.uchicago.edu/?mode=1;id=219}}, located within the region.
A $0.35^{\circ}$ region around the location of 1ES~1215+303 is also excluded.
The exclusion regions are shown in Figure \ref{fig:skymap} as magenta circles.

{}

\par The large area telescope on board the \textit{Fermi} space-based observatory (\lat \cite{FermiLAT}) is a pair-conversion telescope sensitive to gamma rays in the $\sim20\mathrm{~MeV} - > 300 \mathrm{~GeV}$ energy range.
The large FoV of \lat allows for a full survey of the sky every 3 hours.
Data taken by \lat was analyzed using the \texttt{Fermipy} python package \cite{fermipy}, using Science Tools\footnote{\url{https://fermi.gsfc.nasa.gov/ssc/data/analysis/documentation/Cicerone/}} (\texttt{Fermitools v1.0.0}).
``Source'' class events (evclass=128) from both the front and back (evtype=3), with energies between 100 MeV and 300 GeV were considered.
Only photons originating from within a 15$^\circ$ radius of \source were considered and a zenith angle cut of 90$^\circ$ was applied to avoid contamination from the Earth's limb.
A binned-likelihood analysis was performed, where the 3FGL catalog \cite{3FGL} details of nearby sources were included in the model.
The fit was performed with the spectral normalization of sources within 3$^\circ$ of \source allowed to vary and the rest frozen to their 3FGL average.
The normalization of the isotropic and galactic diffuse components were also fit as free parameters.
\source is significantly detected by \lat during this period, with a TS of 441 ($\sqrt{\mathrm{TS}}\sim\sigma$), assuming a power-law model.

\section{Flux Analysis}
\label{sec:flux}

\begin{figure}
    \centering
    \includegraphics[width=\textwidth]{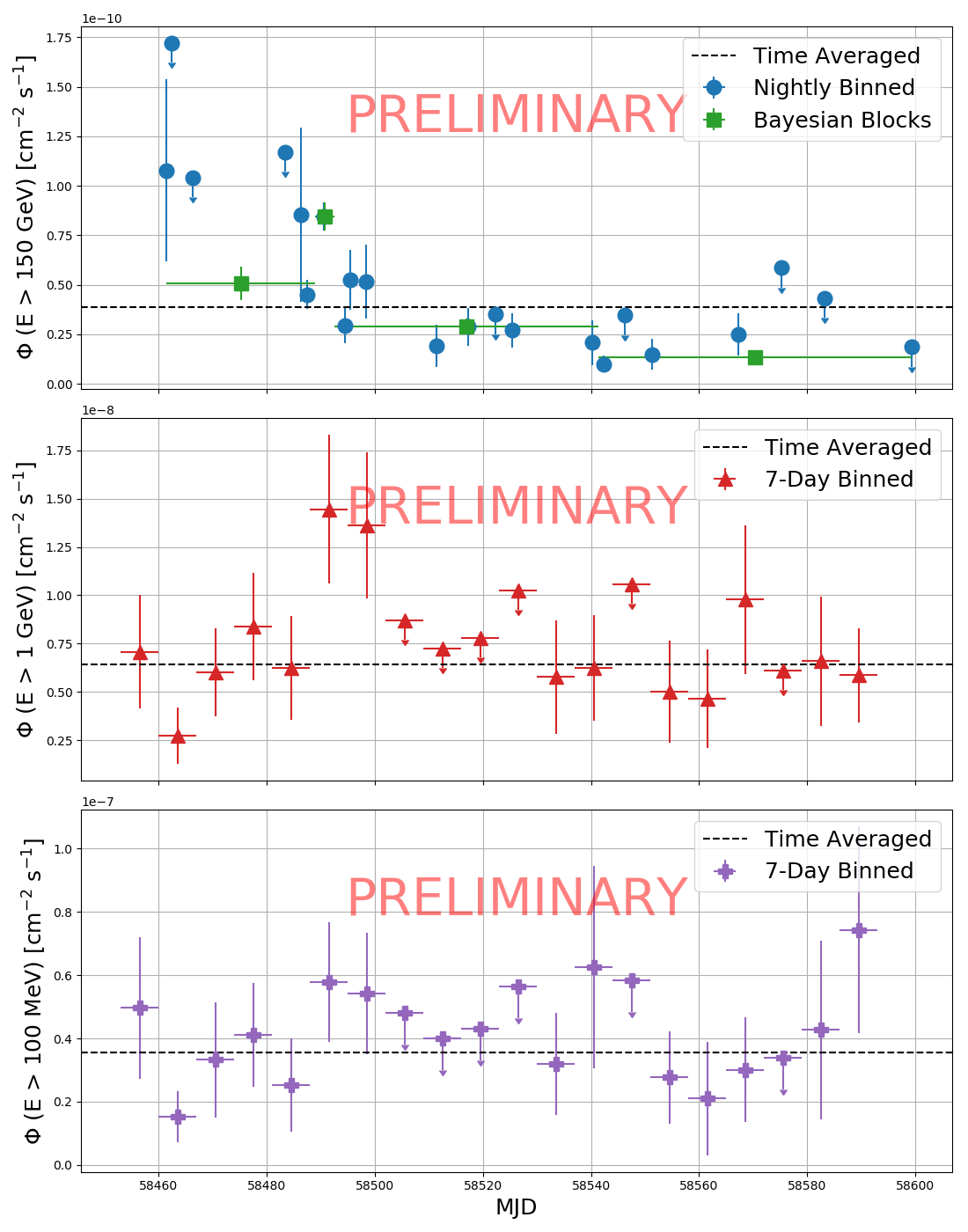}
    \caption{Gamma-ray light curve for \source for the period considered in this analysis.
    The top panel shows the \veritas flux $> 150\mathrm{~GeV}$,
    with the blue circles denoting the nightly-binned flux and the green squares denoting the average flux across the block period.
    The middle panel shows the \lat flux $> 1\mathrm{~GeV}$.
    The bottom panel shows the \lat flux $> 100\mathrm{~MeV}$.
    See text for more details.
    }
    \label{fig:lc}
\end{figure}

The nightly-binned flux ($E>150\mathrm{~GeV}$) obsevered by VERITAS was obtained assuming a power-law model with the spectral index frozen to that of the time-averaged best-fit value of $\Gamma=-3.25$ (see Section \ref{sec:spec} and Table \ref{tab:fit_details}).
The VERITAS flux is shown in the top panel of Figure \ref{fig:lc}.
95\% confidence level (C.L.) upper limits were obtained for bins with less than $2\sigma$ excess significance or $<5$ \texttt{ON}-source events.
The time-averaged flux during this period was determined to be $F(E>150\mathrm{~GeV}) = (3.88 \pm 0.26)\times10^{-11}~\mathrm{~cm^{-2}s^{-1}}$ or 11\% Crab.
To detect significant change points in the VERITAS light curves, a Bayesian-block analysis \cite{scargle} was performed using the \texttt{astropy} python package \cite{astropy}.
In performing this analysis, a false-alarm probability of 0.01 was chosen.
To avoid biasing the results, all data points, regardless of their significance, were included in the analysis.
The time-binned fluxes are shown as green squares in the top panel of Figure \ref{fig:lc}.
The Bayesian-block algorithm picks out a significant day-scale bin on the night of 58490 MJD.
On this night, the flux of \source reaches $F(E>150\mathrm{~GeV}) = (8.44 \pm 0.72)\times10^{-11}~\mathrm{~cm^{-2}s^{-1}}$, or 23\% Crab.
This is $\sim2$ times higher than the time-averaged flux of the campaign.
The spectral analysis of data taken on this night is presented in Section \ref{sec:spec}.

{}

\par Weekly-binned \lat light curves are obtained in the $E>100\mathrm{~MeV}$ and $E>1\mathrm{~GeV}$ energy ranges.
In obtaining the light curve, the fit parameters for other sources within the FoV are frozen to their best-fit values.
The flux $E>100\mathrm{~MeV}$, and $E>1\mathrm{~GeV}$ are shown in the bottom and middle panels of Figure \ref{fig:lc}.
During the campaign, the time-averaged flux in the $E>100\mathrm{~MeV}$, and $E>1\mathrm{~GeV}$ energy ranges were determined to be $(3.55 \pm 0.53)\times10^{-8}~\mathrm{~cm^{-2}s^{-1}}$ and $(6.43 \pm 0.64)\times10^{-9}~\mathrm{~cm^{-2}s^{-1}}$, respectively, and are shown as dashed black lines in the bottom and middle panels.
One notices that the flux above $1\mathrm{~GeV}$  reaches a local maximum which is temporally consistent with the significant day-scale block picked out by the Bayesian-block algorithm.
While the $E>1\mathrm{~GeV}$ flux shows moderate elevation around 58490 MJD, the $E>100\mathrm{~MeV}$ flux shows a level consistent with the campaign mean.
This could be an indication of higher-energy variability, particularly about the inverse-Compton peak (IC-peak) of the SED.
To examine this behaviour, the SED is examined in the next section.

\section{Spectral Analysis}
\label{sec:spec}

The VERTIAS spectral analysis was performed using a forward-folding binned-likelihood method, originally developed for the CAT experiment \cite{piron}.
This method forward-folds the instrument response function with the spectral model to obtain a prediction of the model observed counts, hence allowing for the Poisson based likelihood maximization to be performed.
The time-averaged spectrum was obtained by considering all good-weather data.
The data were divided into bins spaced 0.15 in log-space, and the best-fit power-law model was obtained.
In applying the fit, all bins up to and including the first insignificant bin ($<2\sigma$ excess), were included in the fit.
This aims to the reduce the effect of biasing ones data towards positive fluctuations.
Spectral points are obtained by freezing the spectral shape parameters (i.e. spectral index) and fitting the model across each energy bin, allowing the normalization to vary.
95\% C.L. upper limits are obtained by examining the profile likelihoods of insignificant energy bins, and obtained using a similar method as described by \cite{rolke}.
The details of best-fit time-averaged and ``flare''-night (58490 MJD) are shown in Table \ref{tab:fit_details} and are plotted as green `X's and purple diamonds, respectively, in Figure 3.
Note the points in Figure 3 are deabsorbed for EBL attenuation assuming a \cite{dominguez} EBL model.
The flare-night SED shows a $\sim$2 times increase in the flux normalization.
There is also weak evidence for spectral hardening ($\Delta\Gamma =0.16$), however the change in spectral index is consistent within statistical errors.
\begin{figure}
    \centering
    \includegraphics[width=\textwidth]{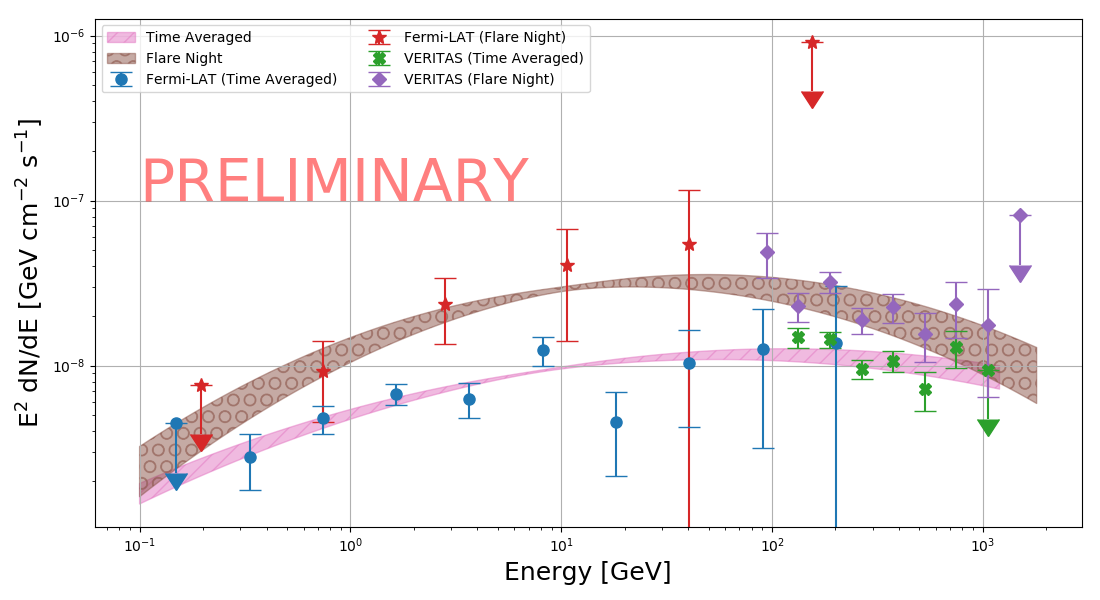}
    \caption{Gamma-ray energy spectrum for \source for the time-averaged data set and the night of the ``flare''.
    The blue circles and red stars denote the time-averaged and flare-night \lat data, respectively.
    The green crosses and purple diamonds denote the time-averaged and flare-night \veritas data, respectively.
    The pink crossed and brown dotted fills denote the best-fit log-parabola models for the time-averaged and flare night data, respectively.
    The data is deabsorbed assuming a \cite{dominguez} EBL model.
    See text for full details.}
    \label{fig:spectra}
\end{figure}{}

\par The \lat energy spectra were obtained for both the time-averaged and the flare-night data.
In taking data for the flare night, all observations taken within $\pm$0.5 days of the VERITAS observations were considered.
This ensures sufficient statistics are available when applying the fit.
The flare-night LAT data set results in a significant detection, with a TS of 74, assuming a power-law model.
The results of the time-averaged and flare-night spectral fits are reported in Table \ref{tab:fit_details} and are shown in Figure \ref{fig:spectra} as blue circles and red stars, respectively.
95\% C.L. upper limits are shown for energy bins with $TS < 9$, corresponding to $\lesssim3\sigma$ detection.
Comparing the best-fit spectral index for two epoch, one sees evidence of spectral hardening $\Delta\Gamma=0.32$, however this is also weak evidence as the change in spectral index is a $<2\sigma$ deviation between the time-averaged and flare-night fits.

\par To characterize the high-energy SED, a joint fit is applied to the \lat and VERITAS data sets.
The data sets were first deabsorbed for EBL absorption assuming a \cite{dominguez} EBL model.
Using the \texttt{Sherpa} python package \cite{sherpa}, the spectral points for both the \lat and VERITAS data were fit with a power-law and log-parabola model.
In applying these fits, the upper limits were excluded from the fitting procedure.
In both the case of the time-averaged and the flare-night data sets, were best-fit by the log-parabola model.
The details of these fits are shown in Table \ref{tab:fit_details} and the 1-$\sigma$ confidence interval on the best fits are shown as shaded regions in Figure \ref{fig:spectra}.
Comparing the best-fit log-parabola model of the time-averaged and flare-night data set show that the spectral index and curvature parameters remain consistent between the different epoch, with flux normalization varying by a factor of $\sim2$. This is consistent with the observed change in flux in the VHE regime.

\begin{table}[]
    \centering
    \resizebox{\textwidth}{!}{
    \begin{tabular}{|l|c|c|c|c|c|}
    \hline
Dataset &	 N &	 $\Gamma$ &	 $\beta$ &	 $E_0$ &	 $\chi^2/NDF$ \\
 & ($\mathrm{cm}^{-2} \mathrm{s}^{-1} \mathrm{TeV}^{-1}$) & & & ($\mathrm{TeV}$) & \\
(1) & (2) & (3) & (4) & (5) & (6) \\
\hline
(Time Averaged) & & & & & \\
\lat  &	 $\left(3.44\pm0.36\right)\times10^{-7}$ &	 $-1.74\pm0.06$ &	 N/A &	 $4.57\times10^{-3}$ &	 N/A \\
VERITAS &	 $\left(2.42\pm0.15\right)\times10^{-10}$ &	 $-3.25\pm0.11$ &	 N/A &	 $2.00\times10^{-1}$ &	 $13.1/5$  ($2.6$) \\
Combined &	 $\left(8.19\pm0.84\right)\times10^{-8}$ &	 $-1.83\pm0.03$ &	 $-0.10\pm0.03$ &	 $1.10\times10^{-2}$ &	 $20.4$/$12$ ($1.7$) \\
\hline
 (Flare Night) & & & & & \\
\lat  &	 $\left(1.16\pm0.35\right)\times10^{-6}$ &	 $-1.42\pm0.17$ &	 N/A &	 $4.57\times10^{-3}$ &	 N/A \\
VERITAS &	 $\left(4.71\pm0.37\right)\times10^{-10}$ &	 $-3.09\pm0.13$ &	 N/A &	 $2.00\times10^{-1}$ &	$9.4/7$ ($1.3$)\\
Combined  &	 $\left(1.69\pm0.41\right)\times10^{-7}$ &	 $-1.86\pm0.07$ &	 $-0.18\pm0.07$ &	 $1.34\times10^{-2}$ &		 $8.4$/$9$ ($0.9$) \\
\hline
    \end{tabular}
    }
    \caption{Fit details for the the different datasets considered. Column (1) shows the data set, Column (2) shows the Normalization at refernce energy given by Column (5). Columns (3) and (4) show the spectral index and curvature parameter where applicable. Column (6) shows the $\chi^2$ per degree of freedom for the fit.
    See text for more details about the fitting procedure.}
    \label{tab:fit_details}
\end{table}{}

\section{Conclusions}
\label{sec:con}
This work represents details of the 2018-2019 monitoring campaign of \source.
During this campaign, enhanced MWL activity was observed from optical to gamma-ray energies.
The analysis of the gamma-ray data during this period reveal a nightly-scale variability, allowing for high-statistics spectra for the time-averaged and flare-night states.
The HE SEDs are obtained in the $100$ MeV to $>1$ TeV energy range, revealing remarkable hard VHE spectra and IC-peaks in the 10s-100s of Gev range.

\par VERITAS will continue to monitor \source and other VHE blazars, as well as initate ToO observations on promising targets triggered by MWL activity.
In the event of enhanced VHE emission, VERITAS will alert multiwavelength partners and the wider astronomy community via astronomers telegrams.

\section*{Acknowledgement}
This research is supported by grants from the U.S. Department of Energy Office of Science, the U.S. National Science Foundation and the Smithsonian Institution, and by NSERC in Canada. This research used resources provided by the Open Science Grid, which is supported by the National Science Foundation and the U.S. Department of Energy's Office of Science, and resources of the National Energy Research Scientific Computing Center (NERSC), a U.S. Department of Energy Office of Science User Facility operated under Contract No. DE-AC02-05CH11231. We acknowledge the excellent work of the technical support staff at the Fred Lawrence Whipple Observatory and at the collaborating institutions in the construction and operation of the instrument.

\bibliographystyle{}

\end{document}